\documentstyle[12pt]{article}
\input{epsf}
\newlength{\extraspace}
\newlength{\extraspaces}
\def\bsklength{.8mm} 

\newcommand{\beq}{\begin{equation}}
\newcommand{\eeq}{\end{equation}}

\newcommand{\beqa}{\begin{eqnarray}}
\newcommand{\eeqa}{\end{eqnarray}}

\newcommand{\newsection}[1]{
\vspace{6mm}
\pagebreak[3]
\addtocounter{section}{1}
\setcounter{equation}{0}
\setcounter{subsection}{0}
\setcounter{figure}{0}
\addcontentsline{toc}{section}{\protect\numberline{\arabic{section}}{#1}}
\noindent{\large \bf \thesection. #1}
\nopagebreak
\medskip
\nopagebreak}

\font\blackboard=msbm10 at 11pt
\font\blackboards=msbm7
\font\blackboardss=msbm5
\newfam\black
\textfont\black=\blackboard
\scriptfont\black=\blackboards
\scriptscriptfont\black=\blackboardss
\def\mathbb#1{{\fam\black\relax#1}}

\def\CD{{\cal D}}

\def\CS{{\cal S}}

\renewcommand{\det}{{\rm det}}

\def\eq{\!\!\!\!=\!\!\!\!}
\def\half{{\textstyle{1\over 2}}}
\def\pa{\partial}

\newcommand{\e}{{\rm e}}

\newcommand{\Tr}{{\rm Tr}}

\def\d{{\rm d}}

\begin{document}
\setcounter{page}{0}
\addtolength{\baselineskip}{\bsklength}
\thispagestyle{empty}
\renewcommand{\thefootnote}{\fnsymbol{footnote}}        

\begin{flushright}
hep-th/9904051\\
\end{flushright}
\vspace{.4cm}

\begin{center}
{\Large
{\bf{Virtual Monopole Geometry and Confinement}}}\\[1.2cm] 
{\rm HoSeong La}
\footnote{e-mail address: hsla@pauli.harvard.edu\\                 
 }\\[3mm]
{\it Lyman Laboratory of Physics\\[1mm]              
Harvard University\\[1mm]
Cambridge, MA 02138, USA} \\[1.5cm]

\vskip1truein

{\sc Preface}\\[1cm]
{\parbox{14cm}{
\addtolength{\baselineskip}{\bsklength}
\noindent
Generalizing the geometry of the gauge covariant variables in Yang-Mills 
theory proposed by Johnson and Haagensen, the 4-d geometry associated with 
a monopole is defined for SU(2). 
There are three relevant geometries: AdS$_2\times S^2$, $R^2\times S^2$
and $H_+\times S^2$, depending on the asymptotic behavior of the torsion.
Using this geometry, the Wilson loop average is computed {\it \`{a} la}
Nambu-Goto action. In case of AdS$_2\times S^2$, it satisfies the area law.

}
}


\end{center}
\noindent
\vfill


\setcounter{page}{1}
\setcounter{section}{0}
\setcounter{equation}{0}
\setcounter{footnote}{0}
\renewcommand{\thefootnote}{\arabic{footnote}}  
\newcounter{xxx}
\setlength{\parskip}{2mm}
\addtolength{\baselineskip}{\bsklength}

\newpage
\pagenumbering{arabic}

\newsection{Introduction}

One of the most intriguing part of Yang-Mills theory is that the fundamental
gauge field variable does not transform covariantly under gauge 
transformations. 
This causes construction of the physical Hilbert space rather complicated,
if not impossible. This in turn makes difficult to investigate the 
nonperturbative aspect of the theory in the strong coupling regime where
perturbation theory fails. Thus we need to look at the theory from 
somewhat different point of view.

One of the proposals made by Haagensen and Johnson is 
to introduce a covariant variable from a geometrical point of view\cite{HJ}.
For example, for SU(2) YM theory a new variable $u_i^a$
can be defined such that a constraint equation relating $u_i^a$ to the 
gauge field $A_i^a$ is
\beq
\label{e1}
\epsilon^{ijk}\left(\pa_j u_k^a +\epsilon^{abc}A_j^b u_k^c \right)=0.
\eeq
For this, the Weyl gauge $A_0^a=0$ is chosen so that the subscript
$i$ runs over the spatial coordinates only. Then $u_i^a$ transforms 
covariantly under gauge transformations.
In a subsequent paper, this condition was generalized to allow a small
nonvanishing rhs\cite{HJL}.
When $u_i^a$ are identified as dreibeins of some geometry, 
this constraint equation is equivalent to nothing but the torsion-free
condition. 
Hence, the following combination can be identified as a metric for 
some geometry associated with YM theory:
\beq
\label{e2}
g_{ij} \equiv u_i^a u_j^a .
\eeq
The local Lorentz symmetry is now SU(2). The hope is rewriting the theory
in terms of $u_i^a$ rather than $A_i^a$ so that the outcome can be manifestly
gauge invariant. Although explicit metrics are 
constructed for instanton or monopole backgrounds\cite{HJL},
the role of such a geometry has not been quite clear so far.

In this paper, we will generalize the above construction including the time
component of the metric for the BPS monopole. The BPS monopole satisfies
YM equation everywhere, yet reduces to
't Hooft-Polyakov monopole asymptotically so that it can clarify
the behavior of the metric better. The BPS monopole is derived usually in
the nonabelian Higgs model context, but such a monopole also exists for 
YM theory without necessarily introducing an extra scalar field. 
The construction is well known as the (anti)self-dual YM equation can be 
reduced to the Bogomol'nyi equation\cite{Ward}.
It is shown in this paper that the resulting 4-d geometries are asymptotically 
${\rm AdS}_2\times S^2$, $R^2\times S^2$, or $H_+\times S^2$, all with a 
nonvanishing torsion. In the ${\rm AdS}_2\times S^2$ case the torsion does not
vanish even in the $r\to\infty$ limit, which distinguishes itself from the 
other two cases. We propose that these are relevant to the
Wilson loop average, as a Nambu-Goto action of some geometry is suspected
to be relevant for the Wilson loop average\cite{Pol}\footnote{An analogous 
conjecture is also used to compute the Wilson loop average in the context of
the AdS/CFT correspondence in string theory\cite{Malda}.}. 
Then we will show how 
it can be related to the confinement in terms of the area law of a Wilson loop.
In fact, we shall find that the relevant geometry for the area law is
${\rm AdS}_2\times S^2$. Other geometries are related to other phases
of YM theory.

\newsection{Virtual Monopole Geometry}

For our purpose of getting 4-d geometry, we will not fix the Weyl gauge. 
Then, let us first define an antisymmetric tensor
field $B^a_{\mu\nu}$ such that
\beq
\label{e1m}
B=du +\omega u + u\omega,
\eeq
where $B = \half B^a_{\mu\nu}T^a \d x^\mu\d x^\nu$, $u=u_\mu^a T^a \d x^\mu$, 
$\omega =\omega_\mu^{\ ab}T^aT^b\d x^\mu$ and 
$\omega_\mu^{\ ab} =- \epsilon^{abc}A_\mu^c$. 
Note that $B$ transforms covariantly under gauge transformations as $u$ does. 
Unlike the 3-d case, here we cannot identify $u$ as a vierbein
and $\omega$ as a connection because $a, b$ indices run over only 3-d
while $\mu$ runs over 4-d.
Nevertheless, we can still construct 
\beq
\label{e2m}
g_{\mu\nu} \equiv u_\mu^a u_\nu^a,
\eeq
which takes the role of 4-d metric. Note that in this construction 
$g_{\mu\nu}$ is not dimensionless because $u_\mu^a$ is not. 
However, at this stage, since there is no
explicit dimensionful parameter, we will not rescale to a dimensionless one.

The corresponding vierbeins, $e_\mu^{\ A}$, $A =(0,a)$, such that
\beq
\label{e3m}
g_{\mu\nu} = e_\mu^{\ A} e_\nu^{\ B} \eta_{AB}, 
\quad \eta_{AB} ={\rm diag} (\pm +++),
\eeq
can be constructed as
\beq
\label{evier}
e_0^{\ 0} \equiv \sqrt{|u_0^a u_0^a|},\quad e_i^{\ a}\equiv u_i^a,
\quad e_i^{\ 0}=0 = e_0^{\ a}.
\eeq
Since $e_0^{\ 0}$ is an SU(2) singlet,
now the ``local Lorentz symmetry'' is enlarged to U(1)$\times$SU(2).
Also the spin connection can be generalized to $\omega_\mu^{\ AB}$ by defining
\beq
\label{espin}
\omega_\mu^{\ 0b} = 0.
\eeq
Then the torsion tensor in this case is given as usual in terms of 
$e_\mu^{\ A}$ and $\omega_\mu^{\ AB}$. In this paper, we are 
mainly interested in the static case so that 
\beq
\label{etors}
T_{ij}^a = B_{ij}^a,\quad
T_{ij}^0 = 0,\quad
T_{0i}^0 =-\pa_i\sqrt{|u_0^a u_0^a|},\quad
T_{0i}^a = \omega_0^{\ ab}u_i^b = B_{0i}^a +\pa_i u_0^a +\omega_i^{\ ab}u_0^b.
\eeq
$T_{0i}^0= -\pa_i e_0^{\ 0}$ already indicates the torsion does not vanish
unless $e_0^{\ 0}$ is constant.

$B$ can be related to the YM field strength as
\beq
\label{ebfs}
dB +\omega B -B\omega = F u - uF.
\eeq

We can construct this geometry for the monopole background as follows.
The relevant BPS monopole solution of self-dual YM equation is
\beqa
\label{emonoa}
A_0^a &=&  {x^a\over r^2}\left({\mu_{{\rm m}}r\over\tanh \mu_{{\rm m}}r} 
-1\right)\\
\label{emonob}
A_i^a &=& \epsilon^{a}_{\ ij}{x^j\over r^2}
\left(1 - {\mu_{{\rm m}}r\over \sinh \mu_{{\rm m}}r}\right),
\eeqa
where $\mu_{{\rm m}}$ is a monopole mass scale. Note that $A_0^a$ takes
the role of the scalar field for the usual monopole solution in the 
Georgi-Glashow model (i.e. SO(3) $\simeq$ SU(2) nonabelian Higgs model). 

We choose a metric whose spatial component is conformally flat:
\beq
\label{e3}
g_{ij} =\mu_{{\rm m}}^2\e^{2\kappa\phi}\delta_{ij},
\eeq
where $\kappa$ is some length scale. The actual magnitude of $\kappa$ is 
quite irrelevant since it can always be rescaled and absorbed into $\phi$.
Thus, we could even choose $\kappa =\mu_{{\rm m}}^{-1}$. 
Now we have introduced in $g_{ij}$ an explicit dimensionful parameter 
coming from the monopole mass scale. Thus the usual dimensionless metric
tensor can be obtained easily by rescaling.
Anyhow, we need such dimensionful parameters to keep track the 
dimensionalities of variables.
Then
\beq
\label{e4}
u_i^a = \mu_{{\rm m}}\e^{\phi/\mu_{{\rm m}}}\delta_i^a.
\eeq
Demanding the spatial part of the torsion vanishes, 
we obtain
\beq
\label{e5}
A_i^a = -{1\over \mu_{{\rm m}}} \epsilon^a_{\ ij}\pa_j\phi.
\eeq

One can easily solve this for the 't Hooft-Polyakov monopole as $r\to\infty$
such that
\beq
\label{e6}
A_i^a = \epsilon^a_{\ ij}{x^j\over r^2},
\eeq
then
\beq
\label{e7}
\phi = -\mu_{{\rm m}}\ln{(\mu_{{\rm m}} r)}.
\eeq

To compute the time component of the metric, it turns out that
we cannot just demand the torsion-free condition for all components of
the torsion. To understand the property of the torsion better, we can
utilize the full BPS monopole to find out the finite $r$ behavior.
In fact, in general the torsion does not have to vanish for finite $r$.
Here we shall first check out the behavior of $B$, which in turn
will let us know about the torsion.
It is reasonable to demand the spatial component of $B$ vanishes 
asymptotically so that
the energy associated with $B$ to be finite, but there is no reason to demand
the same behavior for the time component of $B$. In particular, $B_{0i}^a$
never vanishes for finite $r$.
However, since we do not have any other information for $B$ at this moment,
we will take a compromised position.
In the following we shall demand that the spatial part of the torsion
vanishes everywhere, but leave the time component arbitrary. In this way,
we can still obtain the spatial part of the metric from eq.(\ref{e5}).

Eq.(\ref{e5}) can be solved for the BPS monopole, eq.(\ref{emonob}), to obtain
\beq
\label{emonogeo}
\e^{2\kappa\phi} = {1\over \mu_{{\rm m}}^2 r^2}\tanh^2{\mu_{{\rm m}}r\over 2}.
\eeq
The time component of the metric can be derived from $B_{0i}^a$ as follows.
Let 
\beq
\label{e8}
u_0^a \equiv \mu_{{\rm m}}^2 x^a f(r),
\eeq
then
\beqa
\label{e9}
B_{0i}^a =\!\!\!\!\!\!&&
-\delta_i^a \mu_{{\rm m}}^2f(r) {\mu_{{\rm m}}r\over \sinh \mu_{{\rm m}}r}
-\mu_{{\rm m}}^2x^i x^a \left({1\over r}f'(r) 
+{1\over r^2}
\left(1-{\mu_{{\rm m}}r\over \sinh \mu_{{\rm m}}r}\right)f(r)\right) \cr
&&+\epsilon^{iab}{x^b\over r^3}\tanh {\mu_{{\rm m}}r\over 2}
\left({\mu_{{\rm m}}r\over \tanh \mu_{{\rm m}}r} - 1\right).
\eeqa
If we demand $B_{0i}^a <\infty$ as $r\to\infty$, in general
\beq
\label{e10}
f(r) = c (\mu_{{\rm m}}r)^{-n},\quad (n\geq 0),
\eeq
where $c$ is a constant.
$n$ is chosen to be an integer so that the asymptotic behavior can be 
consistent with the last term of $B_{0i}^a$. 
This choice will also let us avoid later unnecessary
branch cuts when projected onto a two-dimensional surface of Wilson loop
propagation.
We also further demand that $u_0^a < \infty$ as $\mu_{{\rm m}}\to 0$.
This is because eqs.(\ref{emonoa})(\ref{emonob}) are well
behaved in this limit.
As a result, the only allowed $n$ are $n=0,1,2$.

For $n=0$, $B_{0i}^a$ does not vanish asymptotically:
\beq
\label{e11}
n=0:\quad B_{0i}^a \to -c\mu_{{\rm m}}^2 {x^i x^a\over r^2} 
+\epsilon^{iab}\mu_{{\rm m}}{x^b\over r^2} 
\to -c\mu_{{\rm m}}^2 {x^i x^a\over r^2}.
\eeq 
For $n= 1, 2$, 
\beq
\label{e12}
n= 1,2 :\quad B_{0i}^a \to \epsilon^{iab}\mu_{{\rm m}}
{x^b\over r^2} \to 0.
\eeq 
For $n=2$, although the asymptotic behavior of $B$ is 
the same as $n=1$ case, the time component of the resulting metric 
vanishes as $r\to \infty$.
Note that the $\mu_{{\rm m}}$ term behaves the same way as $A_i^a$ 
asymptotically except the prefactor.

The full metric now reads asymptotically
\beq
\label{emet}
\mu_{{\rm m}}^2 ds^2_{{\rm monopole}} 
=c^2\mu_{{\rm m}}^2 \left(\mu_{{\rm m}}r\right)^{2(1-n)} dt^2 
+ {dr^2\over r^2} +d\Omega_2^2, \quad
(n=0,1,2).
\eeq
For $n=0$, it is AdS$_2\times S^2$. For $n=1$, $R^2\times S^2$. 
For $n=2$ the first two terms are the Poincar\'e metric for 
the upper half plane so that the asymptotic topology 
can be identified as $H_+\times S^2$. 

The torsion can be computed accordingly:
\beqa
\label{etora}
T_{0i}^0 &=&|c|(n-1)\mu_{{\rm m}}^2{x^i\over r}{1\over (\mu_{{\rm m}} r)^n},\\
\label{etorb}
T_{0i}^a &=& \epsilon^{iab}{x^b\over r^3}\tanh {\mu_{{\rm m}}r\over 2}
\left({\mu_{{\rm m}}r\over \tanh \mu_{{\rm m}}r} - 1\right)
\to \epsilon^{iab}\mu_{{\rm m}}{x^b\over r^2},
\eeqa
and all other components vanish. Note that in other than $n=0$ case, 
the torsion vanishes in the limit $r\to\infty$.
However, for $n=0$, $T_{0i}^0$ survives even in this limit.

This defines a geometry associated with a monopole in SU(2) YM theory.
$c^2=\pm 1$ is a constant that determines
the signature of the geometry, which can be either Euclidean or Lorentzian
depending on the value of $u_0^a$.
Since this geometry is not that of the spacetime, but associated with the 
dynamical property of the monopole, we call it ``virtual monopole geometry.''
The exact form of the metric for finite $r$ cannot be determined 
at this moment because there is no other information about $B$. 

We can now compute the Wilson loop average in this background.

\newsection{Wilson Loop Average}

It has long been suspected that the Wilson loop average would take the
following form in the leading order\cite{Pol}:
\beq
\label{e15}
W(C)\sim \e^{-\CS_{{\rm NG}}(\Sigma_C)},
\eeq
where $\Sigma_C$ is a minimal area surface bounded by $C$. 
Higher order terms are expected to be related to the 
extrinsic geometry of $\Sigma_C$\cite{Pol,Simonov}.
Intuitively, this is a fairly plausible assumption in the following 
sense. The exponent of the Wilson loop operator can be rewritten as 
an integration over a flat surface bounded by the contour using the Stokes' 
theorem. However, the surface integration is restricted by 
the derivative of the gauge field.
This surface integration is equivalent to a geometric form of 
surface integration over a curved surface in which the classical 
part of the gauge field provides the necessary geometrical information. 
The leading term of this integration is nothing but the Nambu-Goto action
because the latter simply computes the surface area. Any explicit indisputable
proof of this argument does not exist yet for YM theory, but it will
most likely turn out to be true. 

In fact, a rough estimation along the line of the argument in the above 
paragraph shows this is quite reasonable in our case. 
Using the Stokes' theorem, we obtain
\beq
\label{eb1}
\oint_C dx^\mu A_\mu^a = \int_{D_C} d\sigma^{\mu\nu}b_{\mu\nu}^a,
\eeq
where $\sigma^{\mu\nu}$ is a surface element of flat surface $D_C$ bounded
by $C$, and
\beq
\label{eb2}
b_{\mu\nu}^a \equiv \pa_\mu A_\nu^a -\pa_\nu A_\mu^a.
\eeq
Introducing a surface parameter, we can rewrite
\beq
\label{eb3}
\int_{D_C} d\sigma^{\mu\nu}b_{\mu\nu}^a = \int_{\Sigma_C}\! d^2\xi\,
\epsilon^{IJ}\pa_I x^\mu\pa_J x^\nu b_{\mu\nu}^a,
\eeq
where $\Sigma_C$ is now a curved parameter surface.
We can now expand $b$ around the monopole background according to
$A_\mu^a = A_\mu^{(0)a}+\delta A_\mu^a$ such that
\beq
\label{eb4}
b_{\mu\nu}^a = b_{\mu\nu}^{(0)a} + \delta b_{\mu\nu}^a,
\eeq
where
\beqa
\label{eb5a}
b_{ij}^{(0)a} &\eq& 2\epsilon^a_{\ ij}{1\over r^2} 
-2\epsilon^a_{\ jk}{x^k x^i\over r^4} +2\epsilon^a_{\ ik}{x^k x^j\over r^4}
+\cdots,
\\
\label{eb5b}
b_{0i}^{(0)a} &\eq& 
\mu_{{\rm m}}\left(-\delta_i^a {1\over r}+{x^a x^i\over r^3}\right)
+\delta_i^a {1\over r^2}-2{x^a x^i\over r^4}+\cdots .
\eeqa
In the leading order, we obtain the identity\footnote{This is not a gauge 
invariant identification since it involves a specific choice of a background.}
\beq
\label{eb6}
\left(\epsilon^{IJ}\pa_I x^\mu\pa_J x^\nu b_{\mu\nu}^{(0)}\right)^2
= \det\gamma_{IJ},
\eeq
where
\beq
\label{eb7}
\gamma_{IJ}\equiv g_{\mu\nu}\pa_I x^\mu \pa_J x^\nu
\eeq
and $g_{00} =\mu_{{\rm m}}^2,\ g_{ii} = {1\over r^2}$.
This precisely corresponds to $n=1$ case of the metric we constructed before in 
eq.(\ref{emet}).
If we perform the same computation for $\mu_{{\rm m}}\to 0$ limit, we obtain the 
$n=2$ case of the metric. $n=0$ case is slightly more complicated because
$b_{0i}^{(0)a} =0$ for the 't Hooft-Polyakov monopole limit in the Weyl
gauge. However, a careful analysis based on eq.(\ref{emonoa}) in the limit
$\mu_{{\rm m}}\to 0$ first and then $r \to \infty$ leads to
$b_{0i}^{(0)a}\to {x^i x^a\over r^2}$. This leads to $g_{00} = r^2$ and
$g_{ii} = r^{-2}$.
Properly rescaling by $t\to \mu_{{\rm m}}^2 t$, we can obtain the $n=0$ case of
eq.(\ref{emet}). 

This certainly indicates the Wilson loop average is likely of the form 
\beq
\label{e16}
W(C) = \e^{-\int_{\Sigma_C}\sqrt{\gamma} +{\rm h.o.}}\int\CD\delta A_\mu
\e^{iS[\delta A]}
\Tr\,\e^{\int_{D_C} d\sigma^{\mu\nu}\delta b_{\mu\nu}}
\eeq
for a proper action of $\delta A$ derived from the YM action.

Thus, assuming there is no further complication due to the $\delta A$ part, 
here we propose the relevant Nambu-Goto action is the one defined by 
the string of a magnetic flux embedded in the virtual monopole geometry
we constructed:
\beq
\label{e17}
\CS_{{\rm NG}}(\Sigma_C) 
= {1\over 2\pi\alpha'}\int_{\Sigma_C}d^2\xi \sqrt{|\det\gamma_{IJ}|}, 
\eeq
where $g_{\mu\nu}$ is given by eq.(\ref{emet}).
The rational choice of the string tension is
the one associated with the monopole mass scale such that
${1\over 2\pi\alpha'} =\mu_{{\rm m}}^2$ can be chosen.

Fixing the worldsheet coordinates as $\xi^0 = t$ and $\xi^1 =x^1$, we can
compute the Nambu-Goto action over a rectangle in the $(t,x^1)$-plane
with sides $T$ and $R$ to obtain
\beq
\label{e18}
\CS_{{\rm NG}} \sim \mu_{{\rm m}}^{2-n} TV(R).
\eeq
$V(R)$ is equivalent to the potential energy between quarks due to the nature
of the Wilson loop average.
Note that different $n$ values lead to different behaviors of the Wilson loop
average, which suggests that this index is a parameter classifying
different phases of YM theory. 
For given $n =0, 1, 2$, the potential energy between quarks behaves 
for large $R$ like
\beq
\label{epot1}
V(R) \sim{\cases{R,& $n=0$,\cr
          \ln (\mu_{{\rm m}}R),& $n=1$,\cr
          -{1\over R},& $n=2$. }} 
\eeq

The $n=0$ case has the linear potential, hence, it satisfies the area law.
The corresponding metric has asymptotic AdS$_2\times S^2$ topology.
Recall that $n=0$ is the only case that the torsion does not vanish 
in the $r\to\infty$ limit, indicating that the confinement case
has a distinctive geometry compared to other two cases.

The $n=2$ case is the Coulomb phase. 

The third possibility of $n=1$ is 
intriguing, but we suspect that this might be the case of the oblique
confinement\cite{thooft}. This is not actually a far fetched identification.
The monopole solution we have is in fact dyonic for finite $r$. 
This is because the nonvanishing leading order of nonabelian electric
field $E_i^a$ for large $r$ is $-{x^i x^a\over r^4}$, although the contribution
of the 't Hooft-Polyakov monopole limit is zero. Based on the argument
that condensations of these objects cause confinements, it is clear there are
two different possibilities in our case depending on the characteristics 
in which condensations occur. 
Since the dyonic behavior is due to a higher order
contribution, we could identify $n=1$ case as the oblique confinement.
Note that the logarithmic potential is also confining in the sense the 
potential increases asymptotically, even though it does meet the area law
criterion.

\newsection{Final Remarks}

It is shown in this paper that the geometrical structure suggested by
Haagensen and Johnson can be generalized to 4-d and it can take an important
role to prove confinement in YM theory, provided that the Wilson loop
average can be computed in terms of Nambu-Goto action.
The result leads to three possible phases of YM theory, presumably,
confining, oblique confining and Coulomb phases, although the
approach in this paper does not address any dynamical issues of them.

The behavior of different phases are closely related to the property of $B$
field, which is related to the torsion of the 4-d geometry we obtained.
Our result applies only for large $r$, but it is possible to know more about
the finite $r$ region if further information on $B$ is available.
One possible way of incorporating $B$ is to introduce it as the nonabelian 
version of the Kalb-Ramond field. Then it should be possible that 
YM theory can be described by $A$ as well as by $(u, B)$.
In this sense, one can speculate that $B$ could take the role of the dual 
field in the nonabelian case\cite{Pol}.
It will be interesting to see if $B$ is relevant in the string formulation of 
gauge theory.

The result of this paper also suggests that there might be a string theory
on AdS$_2\times S^2$ with a torsion which could be relevant to YM theory
or QCD in 4-d in the spirit of \cite{Maldaone}. The case of string theory
on AdS$_2\times S^2$ without torsion was investigated in \cite{Stro}.

The most important remaining question in this paper is of course if
one can show explicitly that the Wilson loop average in YM theory
is related to the Nambu-Goto action of some string theory. 
In our approach, it is necessary to show that 
$\delta A$ quantum contribution does not spoil the structure.
This will most likely lead to a dual geometrical formulation
of YM theory, presumably a string theory in an extrinsic geometry,
in which $\delta A_\mu$ is properly translated into $\delta g_{\mu\nu}$.
We believe that the geometry provided here is some clue to that.

\bigskip\bigskip
\noindent
{\large\bf{Acknowledgements:}} 
The author thanks C. Vafa for comments and his hospitality. He also thanks
A. Strominger for conversation on \cite{Stro}.



{\renewcommand{\Large}{\large}

}

\end{document}